\documentclass{IEEEtran}
\usepackage{amsfonts,amssymb,amsmath,calc,graphicx,subfigure}
\usepackage{psfrag,array,epsfig,pstricks,multicol,verbatim,moreverb,mathrsfs}
\usepackage{algorithmic, algorithm}
 \usepackage{listings}
\usepackage{color}
\usepackage{cite}

\def\S{{\mathcal{S}}}

\def\Y{{\mathcal{Y}}}

\def\P{\mathrm{Prob}}
\def\Ex{\mathbb{E}}

\begin{document}
\title{Measuring Pulsed Interference in 802.11 Links}
\author{Brad W. Zarikoff, Douglas J. Leith\\Hamilton Institute, NUI Maynooth
\thanks{Supported by Science Foundation Ireland grants 07/IN.1/I901 and 08/SRC/I1403.}}
\maketitle

\begin{abstract}
Wireless 802.11 links operate in unlicensed spectrum and so must accommodate other unlicensed transmitters which generate pulsed interference.   We propose a new approach for detecting the presence of pulsed interference affecting 802.11 links, and for estimating temporal statistics of this interference.  This approach builds on recent work on distinguishing collision losses from noise losses in 802.11 links.   When the intervals between interference pulses are i.i.d., the approach is not confined to estimating the mean and variance of these intervals but can recover the complete probability distribution.   The approach is a transmitter-side technique that provides per-link information and is compatible with standard hardware.    We demonstrate the effectiveness of the proposed approach using extensive experimental measurements.   In addition to applications to monitoring, management and diagnostics, the fundamental information provided by our approach can potentially be used to adapt the frame durations used in a network so as to increase capacity in the presence of pulsed interference.
\end{abstract}

\section{Introduction}
\label{Intro}

\begin{figure}
\centering
\includegraphics[width=0.65\columnwidth]{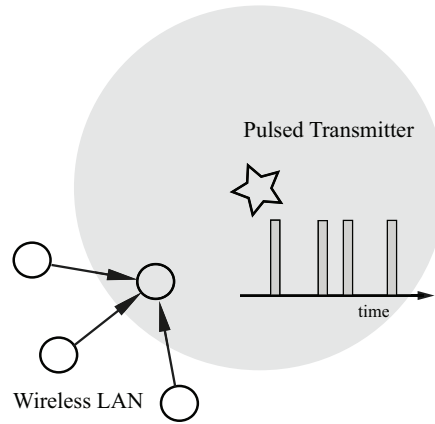}
\caption{Illustrating a WLAN with interfering pulsed transmitter (e.g. 802.11 hidden terminal, Bluetooth device, microwave oven, baby monitor, \emph{etc}) inducing packet loss.}\label{fig:example1}
\end{figure}

Wireless 802.11 links operate in unlicensed spectrum and so must accommodate other unlicensed transmitters.    These transmitters  include not only other 802.11 WLANs but also Bluetooth devices, Zigbee devices, domestic appliances \emph{etc}.  Importantly, the resulting interference is often pulsed in nature.   That is, the interference that consists of a sequence of ``on'' periods (or pulses) during which the interference power is high, interspersed by ``off'' periods where the interference power is lower, illustrated schematically in Fig. \ref{fig:example1}.   The former might be thought of as corresponding to a packet transmission by a hidden terminal and the latter as the idle times between these transmissions.      For this type of interferer, RSSI/SINR measurements are of limited assistance since the SINR measured for one packet may bear little relation to the SINR experienced by other packets.    A further complicating factor is that in 802.11 links frame loss due to collisions is a feature of normal operation in 802.11 WLANs, and thus we need to be careful to distinguish losses due to collisions and losses due to channel impairment.

In this paper we propose a new approach for detecting the presence of pulsed interference affecting 802.11 links and for estimating temporal statistics of this interference under mild assumptions.  Our approach is a transmitter-side technique that provides per-link information and is compatible with standard hardware.  This significantly extends recent work in \cite{txop_technique,measurements} which establishes a MAC/PHY cross-layer technique capable of classifying {lost transmission opportunities} into noise-related losses, collision induced losses, hidden-node losses and of distinguishing these losses from the unfairness caused by exposed nodes and capture effects.

\begin{figure}
\centering
\psfrag{A}[Bc][Bc][0.7]{PHY rate (Mbps)} \psfrag{B}[Bc][Bc][0.7]{Packet Error Rate}
\includegraphics[width=0.7\columnwidth]{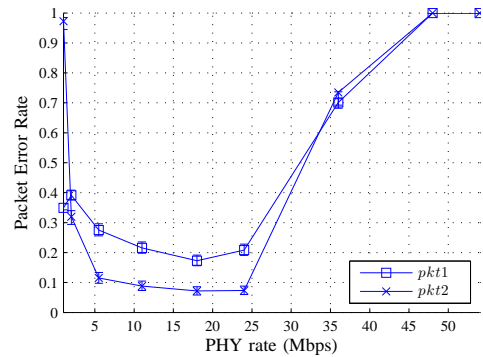}
\caption{Experimental measurements of packet error rate (PER) versus modulation and coding scheme (MCS) for an 802.11 network operating on channel 9 and physically located near an operational microwave oven (MWO).   See Section \ref{sec:mwo} for further details of the experimental setup.   Two curves are shown, one for each fragment of a two packet TXOP burst. Observe that the PER is minimised around 18-24 Mbps and rises at both lower and higher MCS rates due to the pulsed nature of the interference.
}\label{fig:example_uwave}
\end{figure}

Detection and measurement of pulsed interference is particularly topical in view of the trend towards increasingly dense wireless deployments.    In addition to being of interest in their own right for network monitoring, management and diagnostics, our temporal statistic measurements can be used to adapt network parameters so as to significantly increase network capacity in the presence of pulsed interference. This is illustrated in Fig. \ref{fig:example_uwave}, which shows experimental measurements of packet error rate (PER) versus modulation and coding scheme (MCS) for an 802.11 network in the presence of a pulsed microwave oven (MWO) interferer.  Two curves are shown, one for each fragment of a two packet TXOP burst (below we discuss in more detail our interest in using packet pairs).   Observe that the PER is lowest at a PHY rate of 18-24 Mbps -- importantly, the PER rises not only for higher PHY rates, as is to be expected due to the lower resilience to noise at higher rates, but also rises for \emph{lower} PHY rates.    The increase in PER at lower PHY rates is due to the pulsed nature of the interference -- since the frame size in our experiment is fixed, the time taken to transmit a frame increases as the PHY rate is lowered, increasing the likelihood that a frame ``collides'' with an interference burst.   At a PHY rate of 1Mbps, the  frame duration is longer than the maximum interval between interference pulses and, as a result, the PER is close to  100\%.     We discuss this example in more detail in Section \ref{sec:mwo}, but it is clear the appropriate choice of PHY rate can lead to significant throughput gains in such situations.   We briefly note that this type of MAC layer adaptation complements proposed PHY layer interference avoidance techniques such as cognitive radio \cite{cog_rad1}.

\section{Related Work}
\label{Background}

Previous work on estimating 802.11 channel conditions can be classified into three categories. First, \emph{PHY link-level} approaches using SINR and bit-error rate (BER). Second, \emph{MAC approaches} relying on throughput and delay statistics, or frame loss statistics derived from transmitted frames which are not ACKed and/or from signaling messages.  Finally \emph{cross-layer MAC/PHY approaches} that combine information at both MAC and PHY layers.

Most work on PHY layer approaches is based on SINR measurements, e.g. \cite{qiao, Haratcherev,Zhou}. The basic idea is to \emph{a priori} map SINR measures into link quality estimates. However, it is well known that the correlation between SINR and actual packet delivery rate can be weak due to time-varying channel conditions \cite{Aguayo}, pulsed interference being one such example of a time-varying channel. \cite{Rayanchu} considers loss diagnosis by examining the error pattern within a physical-layer symbol, with the aim of exposing statistical differences between collision and weak signal based losses, but does not consider pulsed interference. The cognitive radio literature considers PHY layer techniques for optimising performance in the presence of interference via joint spectral and temporal analysis \cite{cog_rad0}. There are some solutions tailored to the ISM band \cite{cog_rad1}, where customised hardware has been devised with the aim of providing a synchronisation signal based on periodic interference. However, cognitive radio techniques are largely geared towards interference avoidance and make use of non-standard hardware.

MAC approaches make up some of the most popular and earliest rate control algorithms. Techniques such as ARF \cite{ARF}, RBAR \cite{RBAR} and RRAA \cite{RRAA} attempt to use frame transmission successes and failures as a means to indirectly measure channel conditions. However, these techniques cannot distinguish between noise, collision, or hidden noise sources of error. In \cite{Pang}, rate control via loss differentiation is suggested via a modified ARF algorithm; it was shown to greatly improve performance via the inclusion of a NAK signal, but this requires a modification to the 802.11 MAC. Use of RTS/CTS signals has been proposed for distinguishing collisions from channel noise losses, e.g. \cite{cara,dyspan}. However, such approaches can perform poorly in the presence of pulsed interference such as hidden terminals \cite{txop_technique}.

With regard to combined MAC/PHY approaches, the present paper builds upon the packet pair approach proposed in \cite{txop_technique,measurements} for estimating the frame error rates due to collisions, noise and hidden terminals.  See also the closely related work in \cite{PPER_smartrate}.   \cite{txop_technique,measurements,PPER_smartrate} focus on time-invariant channels and do not consider estimation of temporal statistics. \cite{error_id_Zakhor} considers a similar problem to \cite{txop_technique}, but uses channel busy/idle time information.

Some work has been done on packet length adaptation as a means of exploiting a time-varying channel.  \cite{Kim} modifies the Gilbert-Elliott channel model to model bursty channels; however, they do not consider the MAC layer. There are many examples that use MAC frame error information \cite{pktlen-adapt1,pktlen-adapt2,pktlen-adapt3,pktlen-adapt4,pktlen-adapt5}, but  they lack the ability to distinguish between noise and collisions. There has been some recent interesting work on a cross-layer model for packet length adaptation in \cite{Krishnan-infocomm}, which relies on separation between noise errors and collision errors as a means of tuning the packet length and optimising throughput.

\section{Pulsed Interference Temporal Statistics: Non-parametric Estimation}
\label{III}

\subsection{Basic Idea}

\begin{figure}
\centering
\includegraphics[width=0.7\columnwidth]{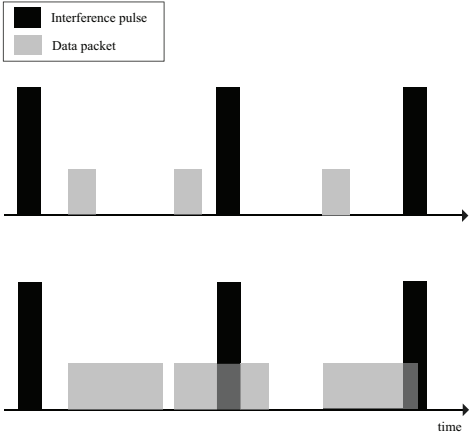}
\caption{Schematic illustrating ``sampling'' of a time-varying channel by data packet transmissions.   Since the data transmissions occupy an interval of time, the sampling is of the channel conditions over that interval, rather than at a single point in time.   As the duration of the data transmissions increases, the chance that a data transmission overlaps with an interference pulse also tends to increase.}\label{fig:pulses}
\end{figure}

We start with the observation that packet transmissions over a time-varying wireless link can be thought of as sampling the channel conditions.   Each sample covers an extended interval of time, equal to the duration $T_D$ of the packet transmission, see Fig. \ref{fig:pulses}.   On a channel with pulsed interference, the frequency with which packet transmissions overlap with interference pulses (and so the level of packet loss) depends on the duration of the packet transmissions relative to the intervals between pulses, and on the durations of the pulses.   For example, it is easy to see that when the packet duration $T_D$ is larger than the maximum time between interference pulses, then every packet transmission overlaps with at least one interference pulse and we can expect to observe a high rate of packet loss.   Conversely, when the packet duration $T_D$ is much smaller than the time between interference pulses, most of the packet transmissions will not encounter an interference pulse and we can expect a much lower rate of packet loss.   Hence, by varying the packet transmit duration and observing the corresponding change in packet loss rate, we can hope to infer information about the timing of the interference pulses.     We can make this intuitive insight more precise as follows.  Assume that the intervals between pulses are i.i.d. so that they are characterised by a probability distribution function.    Then, we will shortly show that the information contained in such packet loss information is sufficient to fully reconstruct this distribution function.   This, somewhat surprising, result has important practical implications.  Namely, that even when the interference pulses are not directly observable (which we expect to usually be the case), we are nevertheless still able to reconstruct key temporal statistics of the interference process from easily measured packet loss statistics.

\subsection{Mathematical Analysis}

We now formalise these claims.  Consider a sequence of interference pulses indexed by $k=0,1,2,...$ and let $T_{k}$ denote the start time of the $k^{th}$ interference pulse with $T_0=0$,  $S_k>0$ denote the duration of the $k^{th}$ pulse and $\Delta_k=T_{k+1}-(T_k+S_k)>0$ be the interval between the end of $k^{th}$ pulse and the start of the $(k+1)^{th}$ pulse.     Defining state vector $X_t:=(t,T_{k(t)},S_{k(t)},\Delta_{k(t)})$, $t\in\mathbb{R}^+$, the sequence $\{X_t\}$ forms a stochastic process with $T_{k+1}=T_{k}+S_k+\Delta_k$, $T_0=0$, $k(t)=\sup \{k:T_k<t\}$.    We assume that the random variables $\Delta_{k},k=1,2,...$ are i.i.d. with finite mean.   Then $\Delta_{k}\stackrel{d}{=}\Delta$, where $\stackrel{d}{=}$ denotes equality in distribution, and let $\P[\Delta\le x]=F(x)$.   Similarly, we assume that the pulse durations $\{S_k\}$ are i.i.d. with finite mean and $S_k\stackrel{d}{=} S$.

Pick a sampling interval $[t-T_D,t]$.   This sampling interval can be thought of as a packet transmission ending at time $t$. Define indicator function $U_{T_D}(X_t)=1$ if interval  $[t-T_D,t]$ does not overlap with any interference pulse, and $U_{T_D}(X_t)=0$ otherwise. That is,
\begin{align}\label{eq:empirical_indicator}
U_{T_D}(X_t) = \left\{
\begin{array}{ll}
1 & t \in [T_{k}+S_{k}+T_D, T_{k+1}) \text{ for some } k\\
0 & \text{otherwise}
\end{array}
\right. .
\end{align}
Suppose we transmit a sequence of packets and let $\{t_j\}$ denote the sequence of times when transmissions finish.   Assume for the moment that (i) a packet is lost whenever it overlaps with an interference pulse and (ii) the intervals between packet transmissions are exponentially randomly distributed and are independent of the interference process.  We will shortly relax these assumptions.   By assumption (i), $U_{T_D}(X_{t_j})$ equals $1$ if the packet transmitted at time $t_j$ is received successfully and $0$ otherwise.     Hence, the empirical estimate of the packet loss rate is
\begin{align}\label{eq:emprical}
\hat{P}_t(T_D) &=1-\frac{1}{N(t)} \sum_{j=1}^{N(t)} U_{T_D}(X_{t_j}),
\end{align}
where $N(t)$ is the number of packets transmitted in interval $[0,t]$.    Provided the packet duration $T_D$ is sufficiently small relative to the mean time between packets, by assumption (ii) the transmit times $\{t_j\}$ effectively possess the Lack of Anticipation property (the number of packet transmissions in any interval $[t,t+u]$, $u\ge 0$, is independent of $\{X_s\}$, $s\le t$ \cite{wolff82}).     When this property holds, by \cite[Theorem 1]{wolff82} we almost surely have
$$\lim_{t\rightarrow\infty}\hat{P}_t(T_D) = \lim_{t\rightarrow\infty}P_t(T_D) =: p(T_D)$$
where
\begin{align*}
P_t(T_D) &=1-\frac{1}{t}\int_{0}^{t} U_{T_D}(X_s)ds.
\end{align*}
That is, the packet loss rate estimator (\ref{eq:emprical}) provides an asymptotically unbiased estimate of the mean value of $U_{T_D}$.

Assumption (i) can be replaced by the weaker requirement that the packet loss rate is higher when a packet transmission overlaps with an interference pulse than when it does not.  We consider this in more detail later, in Section \ref{Est-CM}. Assumption (ii) can be relaxed to any sampling approach that satisfies the Arrivals See Time Averages (ASTA) property, see for example \cite{melamed90, baccelli2009}.

It remains to show that statistic $p(T_D)$ contains useful information about the interference process.  We begin by observing that $Y_t=\sup\{k: T_k\le t\}$ is a renewal process -- since the $\Delta_k$ and $S_k$ are i.i.d., the start times $\{T_k\}$ of the interference pulses are renewal times.    The mean time between renewals is $\Ex[S+\Delta]$.  On each renewal interval $t\in[T_k,T_{k+1}]$ we have that $U_{T_D}(X_t)=1$ for duration $[\Delta_k-T_D]^+$, where $[x]^+$ equals $x$ when $x\ge0$ and $0$ otherwise.   The mean value of $U_{T_D}(X_t)$ over a renewal interval is therefore $\int_{T_D}^{\infty} (x-T_D)dF(x)$ and, by the strong law of large numbers,
 \begin{align*}
p(T_D) &= 1-\frac{1}{\Ex[S+\Delta] }\int_{T_D}^{\infty} (x-T_D)dF(x).
\end{align*}
Since $F(\bullet)$ is a distribution function it is differentiable almost everywhere, and thus so is $p(\bullet)$.  At every point $T_D$ where $p(\bullet)$ is differentiable we have
\begin{align*}
\frac{dp}{dT_D}(T_D)&= \frac{1}{\Ex[S+\Delta]  } \int_{T_D}^{\infty}dF(x) \\
&=\frac{1}{\Ex[S+\Delta]  } \P[\Delta>T_D].
\end{align*}
Provided $p(\bullet)$ is differentiable at $T_D=0$, then
\begin{align*}
\Ex[S+\Delta]  &= \frac{1}{dp(0)/dT_D}
\end{align*}
since $\P[\Delta>0]=1$, and so
\begin{align}
 \P[\Delta>T_D] &= \frac{1}{dp(0)/dT_D}\frac{dp}{dT_D}(T_D). \label{eq:f_T_D}
\end{align}
Hence, knowledge of statistic $p(T_D)$ as a function of $T_D$ is sufficient to allow us to calculate not only the mean time between interference pulses $\Ex[S+\Delta]$, but also the entire distribution function $F(x)=1-\P[\Delta>x]$ of the interference pulse inter-arrival times.

Note that while we can formally differentiate $p(T_D)$, its estimate $\hat{p}(T_D)$  will be noisy and so differentiating $\hat{p}(T_D)$ is not advisable.  The formal differentiation step is merely used to gain insight into the statistical information contained within $p(T_D)$ and there is no need to actually differentiate $\hat{p}(T_D)$ in order to infer characteristics of the interference process (\emph{e.g.} see the examples in the next section).

\subsection{Two Simple Examples}

\begin{figure}%
    \centering
    \subfigure[Periodic interference, period $T_{\Delta}=100$ ms]{%
    \psfrag{A}[Bc][Bc][0.6]{$T_D$ (ms)} \psfrag{B}[Bc][Bc][0.6]{$p(T_D)$} \includegraphics[width=0.45\columnwidth]{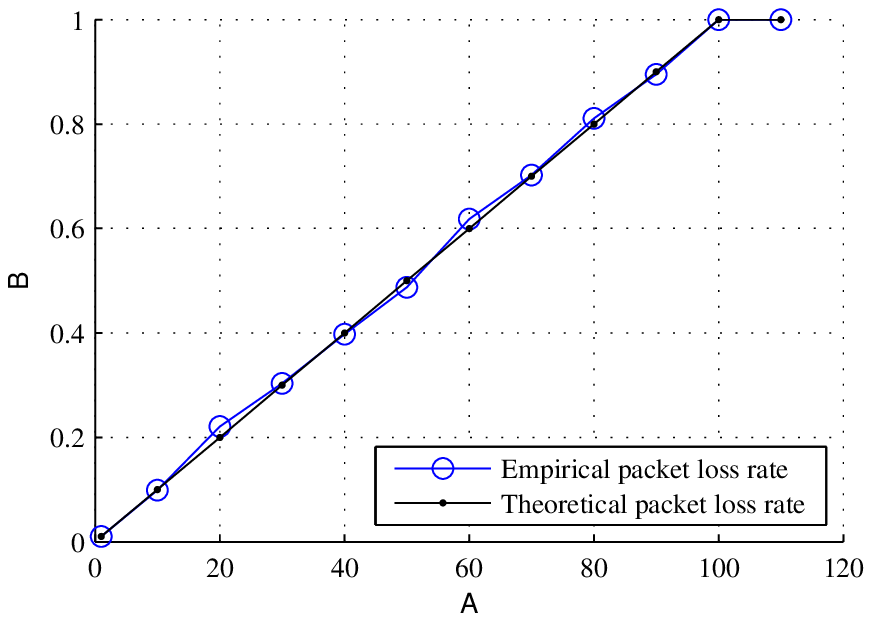}
    }%
    \hspace{8pt}%
    \subfigure[ccdf of $\Delta$ for periodic interference]{\psfrag{A}[Bc][Bc][0.6]{$T_D$ (ms)} \psfrag{B}[Bc][Bc][0.6]{$1-{F}(T_D)$} \includegraphics[width=0.45\columnwidth]{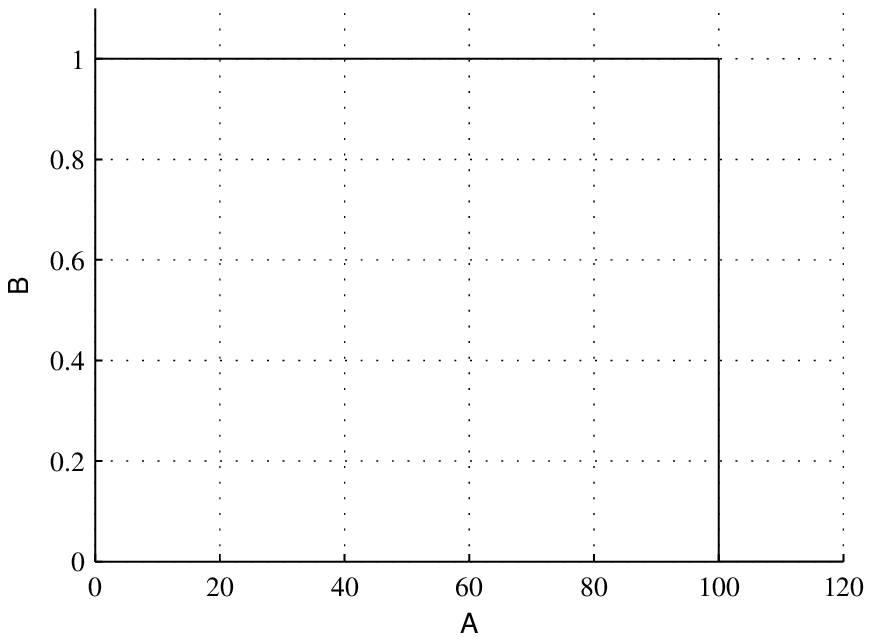}
    }\\
    \subfigure[Poisson interference, mean inter-arrival time $1/\lambda_{\Delta}=10$ ms]{\psfrag{A}[Bc][Bc][0.6]{$T_D$ (ms)} \psfrag{B}[Bc][Bc][0.6]{$p(T_D)$} \includegraphics[width=0.45\columnwidth]{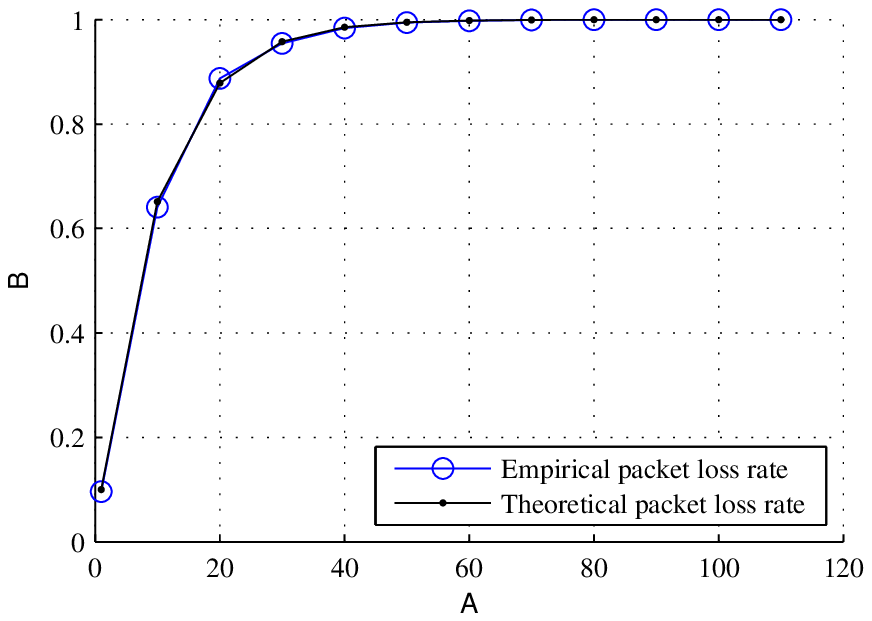}
    }%
    \hspace{8pt}%
    \subfigure[ccdf of $\Delta$ for Poisson interference]{\psfrag{A}[Bc][Bc][0.6]{$T_D$ (ms)} \psfrag{B}[Bc][Bc][0.6]{$1-{F}(T_D)$} \includegraphics[width=0.45\columnwidth]{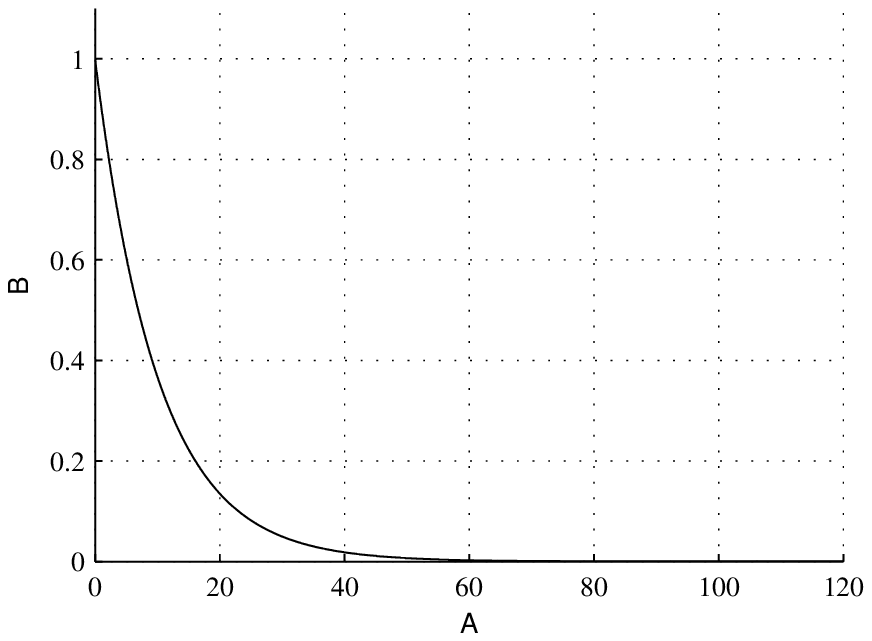}
    }
    \caption{Theory and simulation for periodic and Poisson interference.   Packet transmissions are Poisson with mean rate $\lambda=0.01$.}
    \label{fig:simple}
\end{figure}

We present two simple examples illustrating the use of statistic $p(T_D)$ and for which explicit calculations are straightforward.

\subsubsection{Periodic impulses}
The first example is where the interference consists of periodic impulses with period $T_{\Delta}$ (so $\P(\Delta=T_{\Delta})=1$) and packets are always lost when they overlap with an interference pulse. In this case,
\begin{align*}
p(T_D) &= 1-\frac{1}{\Ex[S+\Delta] }\int_{T_D}^{\infty} (x-T_D)dF(x)\\
&=\left\{
\begin{array}{ll}
\frac{T_D}{T_{\Delta}} & T_D \le T_{\Delta}\\
1 & T_D > T_{\Delta}
\end{array}
\right. .
\end{align*}
That is, $p(T_D)$ is a truncated line with slope $T_{\Delta}$.  Fig. \ref{fig:simple}(a) plots this theory line, along with the measured packet loss rate obtained from simulations. The interference period $T_{\Delta}$ can be directly estimated from the slope of the measured line of packet loss versus $T_D$. The ccdf $1-F(T_D)$ shown in Fig. \ref{fig:simple}(b) can be calculated using (\ref{eq:f_T_D}) or deduced based on the interference period.

\subsubsection{Poisson interference}
The second simple example is where the interference pulses are Poisson impulses, with rate $\lambda_{\Delta}$.   In this case,
\begin{align*}
p(T_D) &= 1-\frac{1}{\Ex[S+\Delta] }\int_{T_D}^{\infty} (x-T_D)dF(x)\\
&=1-\lambda_{\Delta}\int_{T_D}^{\infty} (x-T_D)\lambda_{\Delta}e^{-\lambda_{\Delta}x}dx\\
&=1-e^{-\lambda_{\Delta}T_D}.
\end{align*}
Fig. \ref{fig:simple}(b) shows the corresponding measured packet loss rate obtained from simulations.  Once again, the rate parameter  $\lambda_{\Delta}$ can be directly estimated from the measured curve of packet loss versus $T_D$ (namely from the slope when $p(T_D)$ is plotted on a log scale versus $T_D$).  The ccdf is also shown in Fig. \ref{fig:simple}(d), and calculated as $1-{F}(T_D) = e^{-\lambda_{\Delta} T_D}$.

\subsection{Distinguishing Collision and Interference Losses in 802.11}
\label{IIIc}

The foregoing analysis focuses on packet loss due to interference and ignores other sources of packet loss. As already noted, packet loss due to collisions is part of the proper operation of the 802.11 MAC. In even quite small wireless LANs, the loss rate due to collisions can be significant (\emph{e.g.} in a system with only two users, the collision probability can approach 5\% \cite{nonsat}) and so it is essential to distinguish between packet loss due to collisions and packet loss due to noise/inteference. To achieve this we borrow the packet-pair bursting idea first proposed in \cite{txop_technique}. We make use of the following properties of the 802.11 MAC:
\begin{enumerate}
\item Time is slotted, with well-defined boundaries at which frame transmissions by a station are permitted.
\item The standard data-ACK handshake means that a sender-side analysis can reveal any frame loss.
\item \label{three}Transmissions occurring before a DIFS are protected from collisions.  This is used, for example, to protect ACK transmissions, which are transmitted after a SIFS interval.
\end{enumerate}
Using property \ref{three}, when two frames are sent in a burst with a SIFS between them, the first frame is subject to both collision and noise losses but the second frame is protected from collisions and only suffers from noise/interference losses. Such packet-pair bursts can be generated in a number of ways (\emph{e.g.} using the TXOP functionality in 802.11e/n, or the packet fragmentation functionality available in all flavours of 802.11).

For 802.11 links, we therefore consider sampling the channel using packet pair bursts rather than using single packets. For simplicity we will assume that the duration of both packets is the same and equal to $T_D/2$, although this can be relaxed.   In the remainder of this paper we will often refer to the first packet in a burst as \emph{pkt1}, and the second packet as \emph{pkt2}.    It is important to note that the 802.11 MAC only sends $pkt2$ when an ACK is successfully received for $pkt1$.   To retain the Lack of Anticipation property, when no ACK is received for the first packet we introduce a virtual transmission of the second packet \emph{i.e.} no actual packet is transmitted but the sender still pauses for the time that it would have taken to send the second packet.   In practice this is straightforward to implement by simply adding $T_D/2$ to the interval between packet pairs when an ACK for the first packet is not received.   With this procedure, when the intervals between the completion of one packet pair and the start of the next packet pair form a Poisson process, the packet loss statistics will satisfy the ASTA property.     Assuming that packet collisions occur independently of interference pulses, the packet loss rate for the first packet in the pair $\hat{p}_1(T_D/2)$ is then an estimator for
\begin{align*}
p_1(T_D/2) &= 1-\frac{1-p_c}{\Ex[S+\Delta] }\int_{\frac{T_D}{2}}^{\infty} (x-\frac{T_D}{2})dF(x),
\end{align*}
where $p_c$ is the packet collision probability.     Note that it is difficult to separate out the contribution $p_c$ due to collisions from measurements of $p_1(T_D/2)$, as already discussed.  The second packet in a pair is only transmitted if the first packet was received successfully (per the standard 802.11 TXOP and fragmentation semantics) and so the second packet measurement data is censored.   We therefore have that the packet loss rate for the second packet in the pair $\hat{p}_1(T_D/2)$ is an estimator for
\begin{eqnarray}
p_2(T_D/2) &= 1-\frac{1}{(1-p_1(T_D/2))\Ex[S+\Delta] }\int_{T_D}^{\infty} (x-T_D)dF(x). \nonumber
\end{eqnarray}%
Combining the loss statistics $p_1(T_D/2) $ and $p_2(T_D/2)$ for  the first and second packets, we can recover our desired loss statistic $p(T_D)$ from
\begin{align}
p(T_D)&=1-\left(1-p_2\left(T_D/2\right)\right)\left(1-p_1\left(T_D/2\right)\right),
\label{eq:both_pkts}
\end{align}
and in this way separate out the contribution to packet loss from interference from the contribution due to collisions.

\subsection{Carrier Sense}
\label{IIId}

The 802.11 MAC uses carrier sense to distinguish between busy and idle slots on the wireless medium.  If the energy on the channel is sensed above the carrier-sense threshold, then the PHY\_CCA.indicate(BUSY) signal will be issued by the PHY to indicate to the MAC layer that the channel is busy.   Consequently, when an interference pulse is above the carrier-sense threshold at the transmitter, packet transmissions will not start. Instead, a packet waiting to be transmitted will be queued until the channel is sensed idle (PHY\_CCA.indicate(IDLE)), and then transmitted. This means that the packet transmission times are no longer independent of the interference process and the ASTA property is generally lost. In particular, the packet loss rate is biased and tends to be underestimated since packet transmissions that should have started during an interference pulse (and so likely to have led to a packet loss) are deferred until after the pulse finished (and so much less likely to be lost since the time to the next interference pulse is then maximal).

\begin{figure}
\centering
\subfigure[Packet loss rate versus packet duration $T_D$ with and without carrier sense]{\psfrag{A}[Bc][Bc][0.7]{$T_D$ (ms)} \psfrag{B}[Bc][Bc][0.7]{Packet Loss Rate}
\includegraphics[width=0.7\columnwidth]{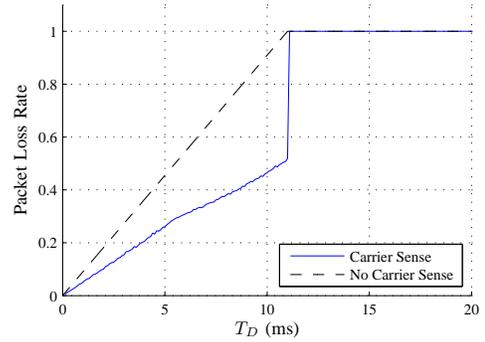}
}
\subfigure[Estimate $\hat{F}(T_D)$ of distribution function $F(T_D)$]{ \psfrag{A}[Bc][Bc][0.7]{$T_D$ (ms)} \psfrag{B}[Bc][Bc][0.7]{$1-\hat{F}(T_D)$}
\includegraphics[width=0.7\columnwidth]{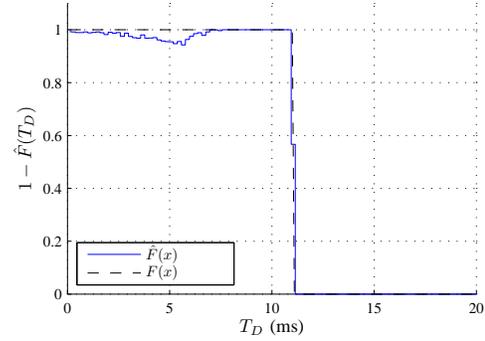}
}
\caption{Simulation example illustrating how the estimation bias introduced by carrier sense can be largely removed using (\ref{eq:simplified_ptilde}).  Periodic interference, similar to the microwave oven interference experimentally measured in Section \ref{sec:mwo} (period ${\Delta}=11$ ms, pulse duration $S=9$ ms).}\label{fig:cs}
\end{figure}

When the duration of the interference pulses is short relative to the time between pulses, then the magnitude of this bias can be expected to be small.    When the interference pulse duration is larger, an approximate compensation for the bias can be carried out as follows. Consider the indicator function
\begin{align*}
\tilde{U}_{T_D}(X_t) = \left\{
\begin{array}{ll}
1 & t \in [T_{k}+T_D, T_{k+1}) \text{ for some } k\\
0 & \text{otherwise}
\end{array}
\right. .
\end{align*}
This modifies (\ref{eq:empirical_indicator}) by lumping the time when the interference pulse is active into the \emph{good} window, roughly capturing the fact that packet transmissions scheduled during a pulse will be deferred until the pulse finishes.   When the interference pulse on and off times are i.i.d., this modified loss statistic is equal with probability one to\newline
\begin{minipage}{\columnwidth}
  \small
\begin{eqnarray}
    \lefteqn{\tilde{p}(T_D) = 1-\frac{1}{\Ex[S+\Delta] }\int_0^{\infty}dG(y)\int_{T_D}^{\infty} (y+x-T_D)dF(x)} \nonumber \\
    & & = 1-\frac{1}{\Ex[S+\Delta] }\left(\Ex[S]F_c(T_D)- \int_{T_D}^{\infty}(x-T_D)dF_c(x)\right)\nonumber\\
    & & = p(T_D) - \epsilon, \label{eq:bias_corrected}
\end{eqnarray}
  \normalsize
\end{minipage}
\vspace{3mm}

\noindent where $F_c(x)=1-F(x)$ is the ccdf, $G(y)=\P[S>y]$ and $\epsilon=\frac{\Ex[S]}{\Ex[S+\Delta] }F(T_D)$ is an approximation to the estimation bias. Using integration by parts and that $\tilde{p}(0) = 1 - \frac{\Ex[S] + \Ex[\Delta]}{\Ex[S+\Delta]}$, (\ref{eq:bias_corrected}) can be rewritten as\newline
\begin{minipage}{\columnwidth}
  \small
\begin{align}
    \tilde{p}(T_D)&= 1 - \frac{\Ex[S]}{\Ex[S]+\Ex[\Delta]} F_c(T_D) + \frac{1}{\Ex[S] + \Ex[\Delta]} \int_{T_D}^{\infty}F_c(x)dx.
    \label{eq:simplified_ptilde}
\end{align}
  \normalsize
\end{minipage}
Assuming that the measured packet loss rate approximates $\tilde{p}(T_D)$, then given measurements of loss rate for a range of $T_D$ values we can solve equation (\ref{eq:simplified_ptilde}) to obtain an estimate for $F(T_D)$ and $\Ex[S]$.    This can be carried out in a number of ways -- one simple approach is to write $F(x)$ as a weighted sum of $\sum_{i=1}^{K} w_i g_i(T_D)$ of orthogonal basis functions $\{g_i(T_D)\}$, and select the weights $\{w_i\}$ and $\Ex[S]$ to minimise the square error between the RHS of (\ref{eq:simplified_ptilde}) and the measurement of the LHS. We illustrate use of this approach in Fig. \ref{fig:cs}, which presents data generated using a simulation with carrier sense and periodic interference. The on-time of the interference pulses is  $S=9$ ms and the time between pulses is ${\Delta}=11$ ms. Fig. \ref{fig:cs}(a) plots the measured packet loss rate versus $T_D$ which is assumed to approximate $\tilde{p}(T_D)$.  Also shown is the loss rate $p(T_D)$ when carrier sense is disabled. The bias $\epsilon$ between $\tilde{p}(T_D)$ and $p(T_D)$ is clearly evident. Using this biased data for $\tilde{p}(T_D)$ and rectangular basis functions $\{g_i(T_D)\}$, solving (\ref{eq:simplified_ptilde}) yields the estimate $\hat{F}(T_D)$ shown in Fig. \ref{fig:cs}(b). It can be seen that $\hat{F}(x)$ accurately estimates the true distribution function $F(T_D)$ (also marked in Fig. \ref{fig:cs}(b)) \emph{i.e.} that we have successfully compensated for the carrier sense bias.    In particular, the sharp transition at 11 ms is accurately estimated.

\section{Experimental Measurements}

In this section, we present experimental measurements demonstrating the power and practical utility of the proposed non-parametric estimation approach. We collected data in two separate measurement campaigns. The first consists of measurements on an 802.11 link affected by interference from a domestic microwave oven (MWO). Such interference is common, and so of considerable practical importance. The second shows measurements from an 802.11 lab testbed, with two transmitting nodes and a number of hidden nodes acting as the pulsed interference source.

\subsection{Hardware and Software}
Asus 700 laptops equipped with Atheros 802.11 a/b/g chipsets (radio 14.2, MAC 8.0, PHY 10.2) were used as client stations, running Debian Lenny 2.6.26 and using a modified Linux Madwifi driver based on 10.5.6 HAL and 0.9.4 driver. A Fujitsu Lifebook P7010 equipped with a Belkin Wireless G card using an Atheros 802.11 a/b/g chipset (AR2417, MAC 15.0, PHY 7.0) was used as the access point, running FreeBSD 8.0 with the RELEASE kernel and using the standard FreeBSD ATH driver. The beacon period is set to the maximum value of 1 s. We disabled the Atheros' Ambient Noise Immunity feature which has been reported to cause unwanted side effects \cite{prop_soln}. Transmission power of the laptops is fixed and antenna diversity is disabled.   In previous work we have taken considerable care to confirm that with this hardware/software setup the wireless stations accurately follow the IEEE 802.11 standard and the packet pair measurement approach is correctly implemented (see \cite{txop_technique, validation, prop_soln} for further details).

A Rohde \& Schwarz FSL-6 spectrum analyser is used to verify that the test channels are unoccupied and also to capture the time-domain traces (see Table \ref{table:SA} for details).

 \begin{table}
    \caption{Spectrum Analyser Details and Setup for Zero Span Measurements.}
    \centering
    \begin{tabular}{|c||c|}
        \hline
        Model & Rohde \& Schwarz FSL6 with optional pre-amp\\
        \hline
        Video BW & 10 MHz \\
        \hline
        Resolution BW & 10 \& 20 MHz \\
        \hline
        Sweep time & 20 ms \\
        \hline
        Antenna & LM Technologies LM254 2.4 GHz dipole \\
        \hline
    \end{tabular}
    \label{table:SA}
\end{table}

\subsection{Microwave Oven Interference}\label{sec:mwo}
\subsubsection{Experimental Setup}

The experimental setup consisted of one client station, the AP and a 700 W microwave oven.   During the experiments, the MWO is operated at maximum power to heat a 2 L bowl of water, and is located approximately 1 m away from the client station and AP; the exact geometry of the setup is not important since the MWO is close enough to the laptops to disrupt communications. The antenna connected to the spectrum analyser is located such that the energy from each RF source is of similar magnitude.

The MWO operates in the 2.4 GHz ISM band, with significant overlap ($>50\%$) with the WiFi 20 MHz channels 6 to 13; this was verified using the spectrum analyser. Our 802.11 experiments used channels 7 and 9 and took place in a room that was cleared for co-channel interference before, during and after each experiment.

The client station transmits packets to the AP with the MTU, FRAG and packet size set to values that ensure that both $pkt1$ and $pkt2$ are of nearly identical duration (the deviation of $T_D/2$ is kept to below 1\%).   The packet duration is adjusted by varying the packet size between 30 and 2110 bytes (yielding $T_D$ from 1.4 ms to 18 ms). These packets are generated using the standard \texttt{ping} command in a bash script. The interval between each set of packet pairs is exponentially distributed with rate $\lambda = 30$ packets per second, and the modulation and coding rate is fixed at 1 Mbps.

\subsubsection{Inferring Interference Statistics From Packet Loss Measurements}

\begin{figure}
    \centering
\subfigure[Measured packet loss rate versus packet duration $T_D$. Confidence intervals based on the Clopper-Pearson method are displayed, but are small enough to be partially obscured by the point markers.]{
    \psfrag{A}[Bc][Bc][0.7]{$\frac{T_D}{2}$ (ms)}
    \psfrag{B}[Bc][Bc][0.7]{\(p(T_D)\)}
    \includegraphics[width=0.7\columnwidth]{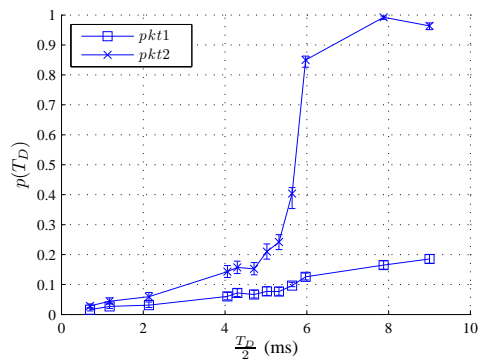}
}
\subfigure[Inter-arrival distribution of interference pulses]{
    \psfrag{A}[Bc][Bc][0.7]{$T_D$ (ms)}
    \psfrag{B}[Bc][Bc][0.7]{$1-\hat{F}(T_D)$}
    \includegraphics[width=0.7\columnwidth]{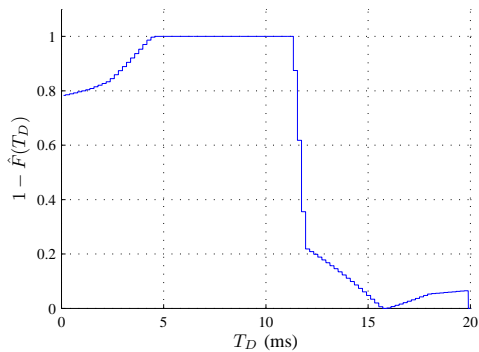}
}
    \caption{Experimental measurements with microwave oven (MWO) interference. Data frames are transmitted at a PHY rate of 1 Mbps rate and the duration \(T_D\) is varied by adjusting the packet size.  Both \(pkt1\) and \(pkt2\) are equal length $\frac{T_D}{2}$.  }
    \label{fig:exp6}
\end{figure}

\begin{figure}
\centering
\subfigure[Packet pair transmitted between two MWO bursts. The y-axis grid is in 2 ms increments. The packet pair is encoded at the 1 MBps 802.11 rate, with both packets having duration 4.36 ms.]{
\includegraphics[width=0.7\columnwidth]{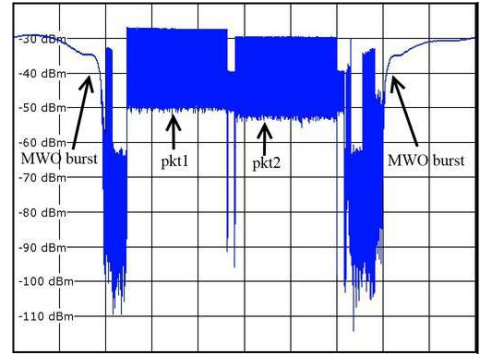}
}
\subfigure[Second packet in a pair suffering a collision with a MWO burst; after the MWO burst has finished and carrier sense indicates the channel is idle, the packet is retransmitted. The y-axis grid is in 2 ms increments.]{
\includegraphics[width=0.7\columnwidth]{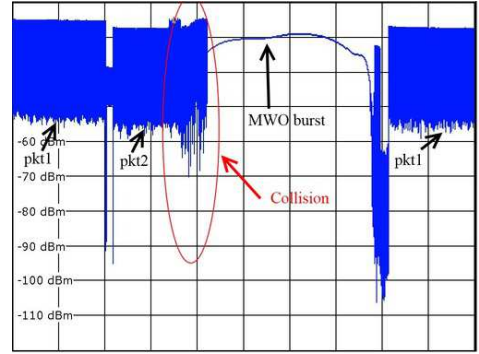}
}
\subfigure[Packet pair and a MWO burst. The y-axis grid is in 2 ms increments. The resolution bandwidth is set to 20 MHz, and thus captures about 99\% of the WLAN signal. The MWO burst has a dip in the middle, which is attributed to frequency instability in the MWO cavity magnetron.]{
    \includegraphics[width=0.7\columnwidth]{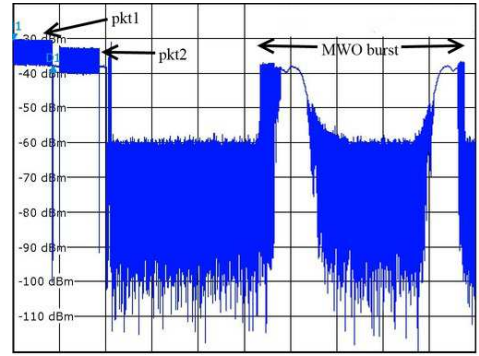}
}
\caption{Spectrum analyser measurements of microwave oven (MWO) interference.}
\label{fig:exp2}
\end{figure}

Fig. \ref{fig:exp6}(a) presents the measured packet loss rate between the client station and the AP versus the packet duration $T_D/2$.   Each point is averaged over more than $10^4$ observed packets.   Using this packet loss data, Fig. \ref{fig:exp6}(b) plots the estimated distribution function $\hat{F}(T_D)$ for interference pulse inter-arrival times.   We use the approach described in Section \ref{IIId} to compensate for the bias introduced by carrier sense at the client station.   It can be seen that $\hat{F}(T_D)$ exhibits a sharp transition around 11 ms, along with some residual probability mass between 11 and 15 ms.    This indicates that the MWO interference is estimated to be approximately periodic with period $\Delta=11$ ms.   We confirm the accuracy of this inference independently using direct spectrum analyser measurements of the MWO interference in the next section, see Fig. \ref{fig:exp2}.

Before proceeding however, it is worth comparing the experimentally measured 802.11 loss data in Fig. \ref{fig:exp6}(a) with the simulation data in Fig. \ref{fig:simple}(a). This comparison highlights the additional complexity introduced by carrier sense and the censoring of second packet loss data.   Nevertheless, our approach is able to successfully disentangle these effects in a principled way and thereby estimate $F(T_D)$.

\subsubsection{Validation}

Fig. \ref{fig:exp2}(a) presents spectrum analyser data showing two interference pulses generated by the MWO. A packet pair transmission by the client station can also be seen, lying between the interference bursts (this particular packet pair transmission is successfully received by the AP, verified by noting the presence of MAC ACKs at the end of each packet). From this and other data, we find that the MWO interference is approximately periodic, with period $T = 1/f = 20\;\mathrm{ms}$ \emph{i.e.} a frequency of $50\;\mathrm{Hz}$, as expected due to the AC circuitry that is driving the MWO.    The profile of the interference bursts is, however, not uniform. Fig. \ref{fig:exp2}(b) shows a measured interference burst of where the interference power is roughly constant over the duration (approximately 9 ms) of the pulse. Fig. \ref{fig:exp2}(c) shows an interference pulse where the interference power dips during the middle of the pulse, so as to effectively create two narrower pulses spaced approximately 4 ms apart. This variation in burst energy profile is attributed to frequency instability of the MWO cavity magnetron, a known effect in MWOs \cite{mwo_int}. Our measurements indicate that the MWO interference consists of pulses with mean interval 11 ms between pulses, with some deviation (Fig. \ref{fig:exp6}(b)). These direct measurements are therefore in good agreement with the estimated distribution function, which was derived indirectly using packet loss measurements.

\subsection{802.11 Network With Hidden Nodes}
\label{homog_ex}

\subsubsection{Experimental Setup}

This test bed consists of a WLAN formed from two client stations and an access point, plus three additional stations configured as hidden nodes. These hidden nodes (HNs) are created by modifying the Madwifi driver such that the carrier sense is disabled (using the technique as detailed in \cite{Anderson2008Commodity}) and setting the NAV to zero for all packets -- this effectively makes the HNs unresponsive to any packets that they decode from the client, or energy that may trigger a physical carrier sense. A script generates \texttt{ping} traffic on the hidden nodes having exponentially distributed intervals between packet transmissions, with a mean interval of 50 ms. The ping packets sent are of duration 4.5 ms (verified via the spectrum analyser). Since the transmissions by each HN are Poisson with intensity $\lambda=20$ packets/s, the aggregate interference is also Poisson and with intensity $\lambda=60$ packets/s.  The experiments used channel 13 of the ISM band, and took place in a room that was cleared for co-channel interference before, during and after the experiments.

\subsubsection{Inferring Interference Statistics From Packet Loss Measurements}

\begin{figure}
 \centering
\subfigure[Measured packet loss rate versus packet duration $T_D$.  Confidence intervals based on the Clopper-Pearson method are displayed, but are small enough to be partially obscured by the point markers.]{
 \psfrag{A}[Bc][Bc][0.7]{$\frac{T_D}{2}$ (ms)}
 \psfrag{B}[Bc][Bc][0.7]{$p(T_D)$}
 \includegraphics[width=0.7\columnwidth]{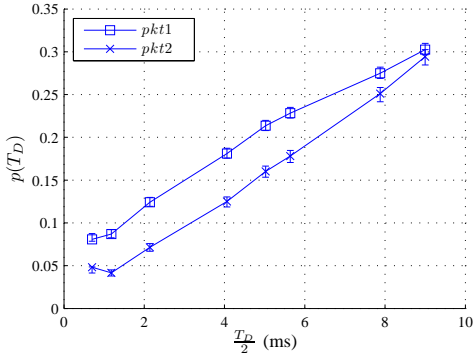}
 }
 \subfigure[Inter-arrival distribution  of interference pulses.]{
 \psfrag{A}[Bc][Bc][0.7]{$T_D$ (ms)}
 \psfrag{B}[Bc][Bc][0.7]{$1-\hat{F}(T_D)$}
 \includegraphics[width=0.7\columnwidth]{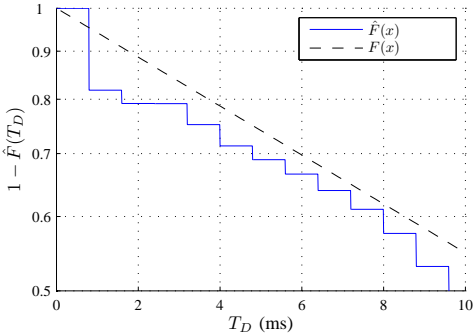}
 }
 \caption{Experimental measurements; primary network has two nodes transmitting to AP, interference network has three hidden nodes. }
 \label{exp_fig1}
\end{figure}

Fig. \ref{exp_fig1}(a) plots the measured packet loss rate in the WLAN versus the packet duration. Note that this loss rate includes a contribution due to collisions between the two client stations in the WLAN and a contribution due to interference from the hidden nodes. Nevertheless, using our packet pair approach we are able to disentangle these two sources of packet loss. Fig. \ref{exp_fig1}(b) plots the resulting distribution of interference pulse inter-arrival times estimated using this packet loss data. The data plotted in Fig. \ref{exp_fig1}(b) is the estimate of $1-F(T_D)$, and is displayed using a logarithmic $y$-axis.   Also plotted in Fig. \ref{exp_fig1}(b) is the theory line $1-F(T_D)=e^{-\lambda T_D}$ corresponding to Poisson distributed interference with rate $\lambda=60$ packets/s.   It can be seen that the estimated data is approximately linear on this log scale, as expected for a Poisson distribution, and that the slope is close to the expected value of $\lambda=60$.    The offset between the Poisson theory line and the estimated line is explained by the presence of a baseline packet loss rate of approximately 5\% in our experimental setup  -- this baseline loss rate is confirmed by separate measurements (not shown here).

\section{Pulsed Interference Temporal Statistics: Parametric Estimation}
\label{Est-CM}

Thus far we have considered estimating the interference distribution function in a non-parametric manner. By making stronger, structural assumptions about the interference process, we can alternatively parameterise the distribution function and our task then becomes one of estimating these model parameters.   A fairly direct trade-off in effort is involved here, which is why it is important to consider both non-parametric and parametric approaches. Namely, we have the bias-variance trade-off whereby non-parametric approaches make only weak assumptions about the interference process, but require more measurement data, whereas parametric approaches make strong assumptions, but require less measurement data for the same estimation accuracy (assuming that the model structure is accurate).

In this section we present a parametric estimation approach for one class of model. The model is related to the two-state Gilbert-Elliot channel model \cite{ge_channel}, which is popular for analysing communication channels with bursty losses, extended to incorporate carrier sensing and the packet transmission process.   Although simple, this model is useful and we demonstrate its effectiveness for estimating hidden terminal interference.  A number of extensions are possible, including to a multi-state interference model \cite{ge_1}, correlated losses \cite{ge_3}, fast fading \cite{ge_2} and so on, but we leave consideration of these extensions to future work.

\subsection{Parametric Packet Loss Model}

\subsubsection{Interference}
\label{Est-CM-2state}

We model pulsed interference as switching randomly between two states, ``good'' ($G$) and ``bad'' ($B$), with exponentially distributed dwell times in each state.    Formally, let $\S=\{G,B\}$ denote the set of interference states,
\begin{align}
    Q&= \left[ \begin{array}{cc}-\lambda_B & \lambda_B\\ \lambda_G  & -\lambda_G\end{array}\right],
\end{align}
and
\begin{align}
    \Pi&= \left[ \begin{array}{cc}0 & 1\\ 1  & 0\end{array}\right].
\end{align}
Let $\Y=\{Y_n, n=0,1,2,...\}$ be a sequence of random variables taking values in $\S$ and representing the evolving state, with
\begin{align}
    \P[Y_{n+1}= j| Y_n=i]&= \Pi_{ij}.
\end{align}
With our the choice of $\Pi$, the $Y_n$ flip back and forth between the $G$ and $B$ states so that $\Y$ is of the form $\{...,G,B,G,B,...\}$.   Let $\{k\}$ index the sub-sequence of $B$ states in $\Y$.   Let $S_k$ denote the dwell time in the $k^{th}$ $B$ state and $\Delta_k$ the dwell time in the following $G$ state.  The dwell times $S_k$ and $\Delta_k$ are independent exponential random variables having, respectively, mean $1/\lambda_B$ and $1/\lambda_G$.   The sequence $T_{k+1}=T_k+S_k+\Delta_k$ is the sequence of jump times at which the interference enters state $B$.

\subsubsection{Packet Transmissions}
\label{Est-CM-PT}

The wireless station performing measurements transmits a sequence of packets to a destination station, with exponentially distributed pauses between transmissions. Similar to the foregoing interference model, we let $\{Tx,Idle\}$ be the two transmitter states, where $Tx$ corresponds to transmission of a packet.  Let $\{V_m,  m=0,1,2,...\}$ denote a sequence of random variables which flip back and forth between the $Tx$ and $Idle$ states. The dwell time in the $Tx$ state is a constant $T_D$, the dwell times in the $Idle$ state are independent exponential random variables with mean $1/\lambda_D$. We index the sub-sequence of $Tx$ states by packet numbers in $\{n\}$, and let $t_n$ denote the time when transmission of packet $n$ starts.

\subsubsection{Carrier Sense}
\label{Est-CM-CS}
The interference state at the packet transmit time $t_{n}$ is $Y_{k(n)}$ where  $k(n)= \sup\{k: T_k\le t_{n}\}$. Let
\begin{align*}
    p_{{cs}} &= \P[t_{n}\in[T_{k(n)},T_{k(n)}+\Delta_{k(n)}]] \\
    &:= \alpha\frac{\lambda_G}{\lambda_G+\lambda_B},
\end{align*}
where $0\le\alpha\le 1$ and $\frac{\lambda_G}{\lambda_G+\lambda_B}$ is the probability than the interference is in state $B$.   In the following, we consider two limiting situations.   Firstly, where the carrier sense threshold lies above the noise level in both interference states, in which case the packet transmission times are decoupled from the interference state and $\alpha=1$.   Secondly,  where the carrier sense threshold lies above the noise level in interference state $G$ but below the noise level in state $B$, in which case $\alpha=0$.

\subsubsection{Packet Loss}
\label{Est-CM-PLP}

Packets are discarded when they fail a checksum test at the receiver. Hence, we treat the channel as an erasure channel. Let $\delta_n$ denote a random variable that takes value $1$ when packet $n$ is erased and value $0$ otherwise. Let $\tilde{S}_{n}$ denote the time that the channel spends in state $B$ during the transmission of packet $n$. In general, we expect that the probability $\P[\delta_n=1]$ that packet $n$ is erased depends on $\tilde{S}_{n}$.  Nevertheless, to streamline the presentation we make the simplifying assumption that $\P[\delta_n=1]=p_B$ whenever $\tilde{S}_{n}>0$ and $\P[\delta_n=1]=p_G$ otherwise, where $p_B$ and $p_G$ are channel packet loss rate parameters in the $B$ and $G$ states respectively. We also assume that packet erasures occur independently, \emph{i.e.} the random variables $\delta_n$, $\delta_m$ are independent for $n\ne m$.

\subsubsection{Packet Error Rate Analysis}
\label{Est-PER1}

\begin{figure}
    \centering
    \psfrag{A}[Bc][Bc][0.7]{\(Idle\)}
    \psfrag{B}[Bc][Bc][0.7]{\(Transmitting\)}
    \psfrag{D}[Bc][Bc][0.7]{\(Loss\)}
    \psfrag{E}[Br][Br][0.7]{\(1-p_{{cs}}\)}
    \psfrag{F}[Bl][Bl][0.7]{\(1-p_i\)}
    \psfrag{G}[Br][Br][0.7]{\(p_{{cs}}\)}
    \psfrag{H}[Br][Br][0.7]{\(p_i\)}
    \psfrag{J}[Bl][Bl][0.7]{\(1\)}
    \includegraphics[width=0.325\textwidth]{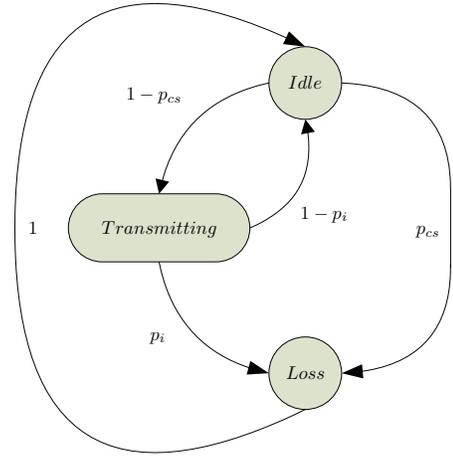}
    \caption{Slotted time Markov chain.}
    \label{fig:traffic_markov_model}
\end{figure}

To determine the packet error rate as a function of the packet transmit duration, we need to analyse two coupled stochastic processes, namely the channel and transmission processes. The joint process takes state values in $\{G,B\}\times\{Idle,Tx\}$.    Since our interest is in counting the frequency of packet losses, observe that we can lump the $(Idle,G)$ and $(Idle,B)$ states together, since we know that no packet loss can occur in these $(Idle,\bullet)$ states.   Also, when the system first enters state $(Tx,B)$, then a packet loss occurs and we do not need to keep count of the number of subsequent transitions between $(Tx,G)$ and $(Tx,B)$.   We can therefore partition time into slots, with each slot being of three possible types: $Idle$ (corresponding to the lumped $(Idle,\bullet)$ states), $Loss$ (corresponding to lumping of states $(Tx,G)$ and $(Tx,B)$ after the first transition from $(Tx,G)$ to $(Tx,B)$) and $Transmitting$ (corresponding to a dwell time in state $(Tx,G)$).   The transitions between these slots are as shown in Fig. \ref{fig:traffic_markov_model} and Table \ref{table:1}.

\begin{table*}
    \renewcommand{\arraystretch}{1.3}
    \caption{Markov model state transitions. }
    \label{table:1}
    \centering
    \begin{tabular}{|c|c|c|}
        \hline
        $Idle \rightarrow Transmitting$ &\emph{(start Tx, interference in state G)}: &
        $1-p_{{cs}}$ \\
        \hline
        $Idle\rightarrow Loss$ &\emph{(start Tx, interference in state B)}: &    $p_{{cs}}$ \\
        \hline
        $Transmitting\rightarrow Idle$ &\emph{(interference in state G throughout Tx)}: &   $1-p_i =\exp(-\lambda_BT_D )$ \\
        \hline
        $Transmitting\rightarrow Loss$ &\emph{(interference enters state B during Tx)}: &   $p_i=1-\exp(-\lambda_B T_D )$ \\
        \hline
        $Loss\rightarrow Idle$ &\emph{(Tx of damaged packet ends)}: & $1$ \\
        \hline
    \end{tabular}
\end{table*}

The transition matrix $P$ of this slotted time Markov chain is:
\begin{equation}
    \mathbf{P} = \left[ \begin{array}{ccc}
        0 & 1-p_{{cs}} & p_{{cs}} \\
        1-p_i(T_D) & 0 & p_i(T_D) \\
        1 & 0 & 0
        \end{array} \right],
\end{equation}
where $1-p_i(T_D)=\exp(-\lambda_B T_D )$.  The stationary state distribution satisfies $\pi = \pi \mathbf{P}$, where $\pi_1 = \P[Idle]$, $\pi_2 = \P[Transmitting]$, and $\pi_3 = \P[Loss]$. Solving yields,
\begin{align*}
    \pi^T &=  \frac{1}{2+p_i(T_D)\left(1-p_{{cs}}\right)} \left[
    \begin{array}{c}
        1 \\
        1-p_{{cs}} \\
        \left(1-p_{{cs}}\right) p_i(T_D) + p_{{cs}}
    \end{array}
    \right]. \nonumber
\end{align*}
The packet error probability for the first packet in a pair is
  \begin{minipage}{\columnwidth}
    \small
\begin{eqnarray}
    \lefteqn{p_1(T_D) = \frac{(1-p_i(T_D))\pi_2p_G + p_i(T_D) \pi_2 p_B + p_{{cs}} \pi_1 p_B}{(1-p_i(T_D)) \pi_2 + p_i(T_D) \pi_2 + p_{{cs}} \pi_1}} \nonumber \\
    && = (1-p_i(T_D))(1-p_{{cs}})p_G + \left(p_i(T_D) \left(1-p_{{cs}}\right)+p_{{cs}}\right) p_B \nonumber\\
    &&=:G_1(T_D,\lambda_B, p_B, p_G, p_{{cs}}).\label{eq:Pe1}
\end{eqnarray}
   \normalsize
  \end{minipage}

\vspace{3mm} \noindent The first term in the expression for $p_1(T_D)$ corresponds to the event where the interference stays in state $G$ throughout a packet transmission and a packet loss occurs. The second term corresponds to the event that a packet transmission starts with the interference in state $G$, but the interference changes to state $B$ during the course of the transmission and a packet loss occurs. The third term corresponds to the event that a packet transmission starts with the interference in state $B$ and a packet loss occurs.

Conditioned on the first packet transmission being successful, the packet error probability for the second packet in a pair is
\begin{eqnarray}
    p_2(T_D) &=& (1-p_i(T_D))\frac{\lambda_B}{\lambda_B + \lambda_G} p_G \nonumber\\
    & & + \left( 1 - (1 - p_i(T_D)) \frac{\lambda_B}{\lambda_B + \lambda_G}  \right) p_B \nonumber \\
    & =: & G_2(T_D,\lambda_B, p_B, p_G, p_{{cs}}), \label{eq:Pe2}
\end{eqnarray}
where the $\frac{\lambda_B}{\lambda_B + \lambda_G} $ factor accounts for the event that the interference is in the $B$ state upon starting transmission of $pkt2$.

\subsection{Model parameters}
Equations (\ref{eq:Pe1}) and (\ref{eq:Pe2}) together form a parametric model of the packet pair loss process, which is described by parameters $\lambda_B$, $p_B$, $p_G$ and $p_{{cs}}$.

\begin{figure}
 \centering
 \psfrag{A}[Bc][Bc][0.7]{\(T_D\) (ms)}
 \psfrag{B}[Bc][Bc][0.7]{\(p(T_D)\)}
 \psfrag{C}[Bl][Bl][0.7]{\(\lambda_B=100\)} \psfrag{D}[Br][Br][0.7]{\(\lambda_B=1000\)}
 \includegraphics[width=0.7\columnwidth]{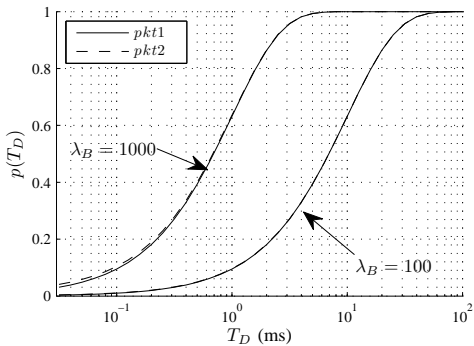}
 \caption{Packet error rate versus packet duration $T_D$;  \( \lambda_D = 30\), variable \(\lambda_B\), \(p_G = 0\), \(p_B = 1\), \(p_{{cs}} = 0\), $T_{SIFS}=10\;\mu$s. }
 \label{fig1}
\end{figure}

\begin{figure}
 \centering
 \psfrag{A}[Bc][Bc][0.7]{\(T_D\) (ms)}
 \psfrag{B}[Bc][Bc][0.7]{\(p(T_D)\)}
 \psfrag{D}[Br][Br][0.7]{\(p_B = 0.9\)}
 \psfrag{C}[Bl][Bl][0.7]{\(p_B = 0.7\)}
 \includegraphics[width=0.7\columnwidth]{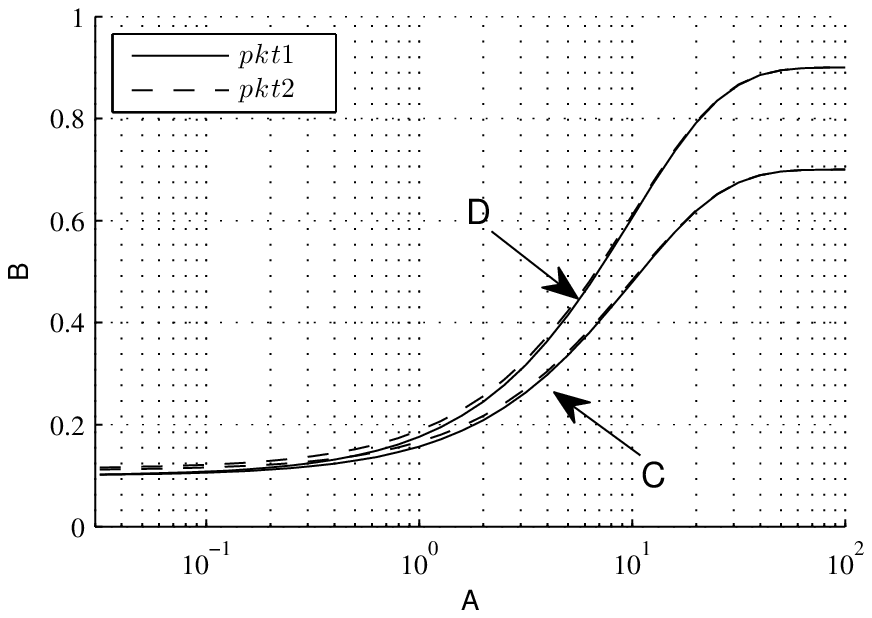}
 \caption{Packet error rate versus packet duration $T_D$; \( \lambda_D = 30\), \( \lambda_B = 100 \), \(p_G=0.1\), variable \(p_B\), \(p_{{cs}} = 0\), $T_{SIFS}=10\;\mu$s.}
 \label{fig2}
\end{figure}

\begin{figure}
 \centering
 \psfrag{A}[Bc][Bc][0.7]{\(T_D\) (ms)}
 \psfrag{B}[Bc][Bc][0.7]{\(p(T_D)\)}
 \psfrag{C}[Bl][Bl][0.7]{\(\alpha = 0\)}
 \psfrag{D}[Bl][Bl][0.7]{\(\alpha = 0.2\)}
 \psfrag{E}[Bl][Bl][0.7]{\(\alpha = 0.8\)}
 \psfrag{F}[Bl][Bl][0.7]{\(\alpha = 1\)}
 \includegraphics[width=0.7\columnwidth]{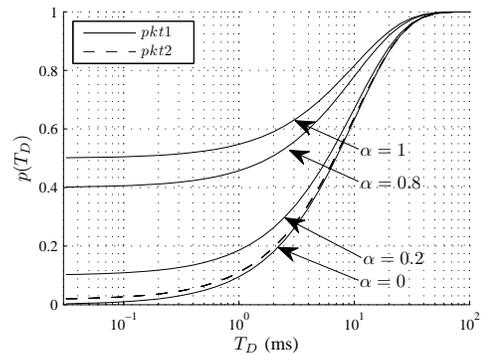}
 \caption{Packet error rate versus packet duration $T_D$; \( \lambda_D = 30\), \( \lambda_B = 100 \), \(p_G = 0\), \(p_B = 0\), variable \(p_{{cs}}\) (by varying \(\alpha\)), $T_{SIFS}=10\;\mu$s.}
 \label{fig3}
\end{figure}

Before proceeding, we briefly illustrate how the model parameters $\lambda_B$, $p_B$, $p_G$ and $p_{{cs}}$ affect the observed packet loss versus $T_D$ curves.   Our aim is to (i) illustrate the types of loss curves that the model is able to capture and (ii) gain some intuitive insight into the role of the various model parameters.  Fig. \ref{fig1} shows the impact of $\lambda_B$, which produces a horizontal shift in the loss curves.  Fig. \ref{fig2}  shows the impact of $p_B$, which determines the right-hand asymptote of the loss curves.    Fig. \ref{fig3} shows the impact of the carrier sense parameter $p_{{cs}}$ (by varying $\alpha$), which produces a vertical shift in the left-hand asymptote. Although not shown, the impact of $p_G$ also produces a vertical shift in the left-hand asymptote.

\subsection{Maximum Likelihood Parameter Estimation}
\label{Est-SD}

Our objective is to estimate the model parameters $\lambda_B$, $p_B$, $p_G$ and $p_{{cs}}$ from measurements of packet loss.   The empirical estimators for  loss probabilities $p_1(T_D)$ and $p_2(T_D)$ are
\begin{align*}
    \hat{p}_1(T_D) &= \frac{1}{N_1}\sum_{n=1}^{N_1} \delta_n^1 &
    \hat{p}_2(T_D)  = \frac{1}{N_2}\sum_{n=1}^{N_2} \delta_n^2,
\end{align*}
where $N_1$ is the number of first packets, $N_2$ the number of second packets, $\delta_n^1$ is the indicator function that equals $1$ when the $n^{th}$ first packet is lost and $0$ otherwise, and similarly $\delta_n^2$ for second packets.     Collecting packet loss measurements for a sequence of packet durations $T_{D_1}, T_{D_2}, ...$ and stacking the corresponding loss probability estimates we have
\begin{align}
    \left[
        \begin{array}{lr}
            \hat{p}_1(T_{D_1})\\ \hat{p}_2(T_{D_1}) \\
            \hat{p}_1(T_{D_2})\\\hat{p}_2(T_{D_2})\\
            \vdots
        \end{array}
    \right]
    =
    \left[
        \begin{array}{lr}
            G_1(T_{D_1},\lambda_B, p_B, p_G, p_{{cs}})\\ G_2(T_{D_1},\lambda_B, p_B, p_G, p_{{cs}})\\
            G_1(T_{D_2},\lambda_B, p_B, p_G, p_{{cs}})\\ G_2(T_{D_2},\lambda_B, p_B, p_G, p_{{cs}})\\
            \vdots
        \end{array}
    \right]
    +\eta, \label{eq:ls}
\end{align}
where $\eta$ denotes the estimation error in the packet loss estimates.    For $N_1$, $N_2$ sufficiently large,  the estimation error $\eta$ is close to being Gaussian distributed.     The maximum likelihood estimates for parameters $\lambda_B$, $p_B$, $p_G$ and $p_{{cs}}$ are then the values that minimise the square error between the LHS and RHS in (\ref{eq:ls}).

\subsection{Experimental Measurements}
\label{PE}
\subsubsection{Experimental Setup}
We revisit the WLAN experimental setup discussed in Section \ref{homog_ex}, but now change the setup slightly so that only a single wireless client (rather than two clients) transmits in the WLAN.   This change is introduced because, for simplicity, we have not included packet collisions in our parametric model.

\subsubsection{Packet Loss Measurements}
\begin{figure}
    \centering
     \psfrag{A}[Bc][Bc][0.7]{\(T_D\) (ms)}
     \psfrag{B}[Bc][Bc][0.7]{\(p(T_D)\)}
    \psfrag{D}[Br][Br][0.7]{3 int}
    \psfrag{C}[Bl][Bl][0.7]{1 int}
    \includegraphics[width=0.7\columnwidth]{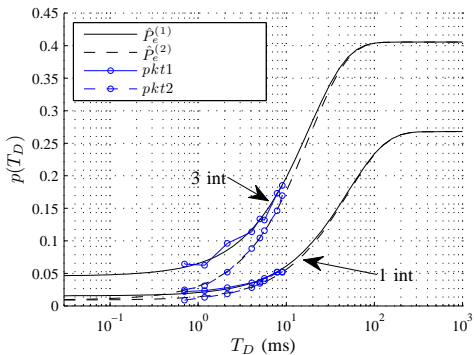}
    \caption{Experimental measurements and model fit for WLAN with hidden node interference. Data points are for experiments using 1 and 3 interferers, with each interferer having a packet transmission rate of \(\lambda_B = 20\). Initial values for the parameter estimator are $\hat{\lambda}_B = 20$, $\hat{p}_{cs} = 0$, $p_G = 0$, and $p_B = 0.5$. Model parameters are given in Table \ref{table1}.}
    \label{fig:8}
\end{figure}

Fig. \ref{fig:8} shows the measured packet loss rate versus the packet duration $T_D$.  Note that the range of packet durations that we can use is constrained by the maximum 802.11 frame size of 2272 B to lie in the interval 1.4 ms to 18 ms.    Two sets of results are shown, for one and for three hidden nodes active.   Each experimental point is calculated as the average of more than $6\times 10^5$ packet transmissions.    Also shown are the maximum likelihood fits to this data using parametric model (\ref{eq:Pe1}) and (\ref{eq:Pe2}); the corresponding model parameter estimates are given in Table \ref{table1}, obtained using an interior-point solver.

\begin{table}
    \centering
    \begin{tabular}{|c||c|c|c|c|}
        \hline
          Number of interferers & \(\hat{\lambda}_B\) & \(\hat{p}_{{cs}}\) & \(\hat{p}_G\) & \(\hat{p}_B\) \\
        \hline \hline
         1 & 19.9932 & 0.0286 & 0.0080 & 0.2678 \\
        \hline
         3 & 54.7173 & 0.1011 & 0.0055 & 0.4055 \\
        \hline
    \end{tabular}
    \caption{Details of the maximum likelihood parameter estimates for measurement data in Fig. \ref{fig:8}}
    \label{table1}
\end{table}

\subsubsection{Validation}
The hidden node interferers each make transmissions with exponentially distributed idle time between packets so that the mean transmit rate is 20 packets/s.   When one interferer is active, we expect $\lambda_B=20$ and when three interferers are active we expect $\lambda_B=60$.   It can be seen from Table \ref{table1} that the model estimates are close to these predictions.   Measurements taken with no hidden node interferers active indicate that the baseline packet loss rate is less than 1\% and it can be seen from Table \ref{table1} that the model estimate for $p_G$ is in good agreement with this.    While it is difficult to similarly validate the estimates for parameters $p_{{cs}}$ and $p_B$, we note that the estimated values are very reasonable.

\subsubsection{Parametric vs Nonparametric Estimation}
\begin{figure}
    \centering
    \psfrag{A}[Bc][Bc][0.7]{\(N\)}
    \psfrag{B}[Bc][Bc][0.7]{\(\max_x | \hat{F}_\infty(x) - \hat{F}_N(x)|\)}
    \includegraphics[width=0.7\columnwidth]{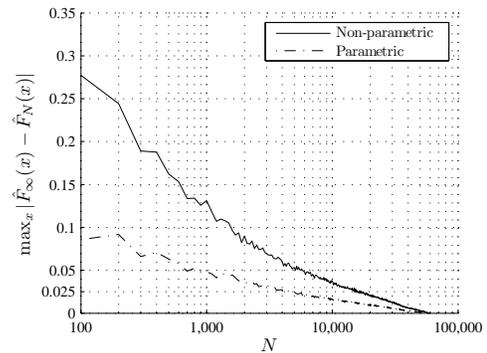}
    \caption{Convergence of estimates of $F(x)$ versus the number of packets observed.    $\hat{F}_N(x)$ denotes the estimate using $N$ packet observations and $\hat{F}_\infty(x)$ denotes the estimate obtained using the full measurement trace.  For each $N$, we take 100 random subsamples of $N$ packets from the full measurement trace, calculate $\max_x|\hat{F}_\infty(x)-\hat{F}_N(x)|$ for each subsample, and average this value over the 100 subsamples to obtain the curves shown.  Data is shown for both parametric and non-parametric estimates.  The data set used is from the three interferer experiment, see Fig. \ref{fig:8}. }
    \label{fig:converge}
\end{figure}

A parametric model makes strong structural assumptions that allow the loss curves to be parameterised using a small number of parameters.    Since there are fewer parameters, we expect to be able to estimate their values with less data, but at the cost of introducing a bias if the structural assumptions turn out to be incorrect.    Fig. \ref{fig:converge} plots $\max_x |\hat{F}_\infty(x)-\hat{F}_N(x)|$ versus the number of observed packets $N$ for both the parametric and non-parametric approaches, where $\hat{F}_N(x)$ is the estimate of $F(x)$ obtained using $N$ observations and $\hat{F}_\infty(x)$ is the estimate using all $6\times 10^5$ observations. For the parametric model, the parameter estimates are fed back into the model equations (\ref{eq:Pe1}) and (\ref{eq:Pe2}), and the resulting parameterised $P_e$ curves are used to calculate $\hat{F}_N(x)$. This provides a rough indication of how estimates converge as the amount of data is increased.   It can be seen that the parametric solution converges to within 5\% of the asymptotic estimate after $N=900$ packets and to within 2.5\% after $N=4000$ packets, while the non-parametric solution requires $N=6000$ and $N=20000$ packets, respectively, to achieve the same level of estimation accuracy.

\subsubsection{Discussion}

\begin{figure}
    \centering
    \includegraphics[width=0.7\columnwidth]{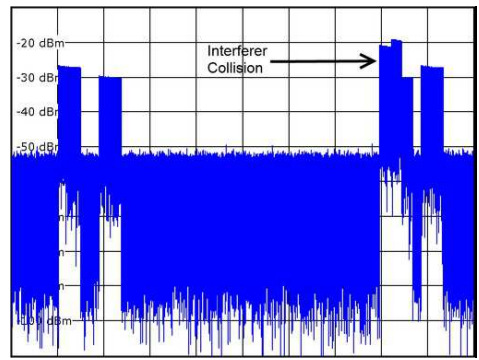}
    \caption{Spectrum analyser snapshot of hidden terminal interferers in time. The y-axis grid is in 2 ms increments. Interferer burst durations are fixed at 4.5 ms, with arrivals at 10, 19, 80, 83 and 89 ms. Since each interferer has a different path to the spectrum analyser antenna, the pulses are at different power levels. The third and fourth pulses collide, resulting in a stepped feature.}
    \label{fig:8a}
\end{figure}

It is interesting to note that, despite its simplicity, the parametric model used here is remarkably effective at capturing the behaviour in a complex physical environment. For example, the model ignores the fact that the interference power will depend on the \emph{number} of hidden node transmissions taking place at the same time. This effect can be seen in the spectrum analyser measurements in Fig. \ref{fig:8a}, where overlapping transmissions by interferers leads to a stepped interference pulse profile.   The model also assumes that the duration of interference pulses is exponentially distributed, but this will not be the case in our experimental setup. More complex parametric models are also possible, and in particular can leverage the wealth of research on bursty communications channels, but we leave this to future work.

\section{Conclusions}

In this paper we propose a new approach for detecting the presence of pulsed interference affecting 802.11 links, and for estimating temporal statistics of this interference. Our approach is a transmitter-side technique that provides per-link information and is compatible with standard hardware.  This significantly extends recent work in \cite{txop_technique,measurements} which establishes a MAC/PHY cross-layer technique capable of classifying {lost transmission opportunities} into noise-related losses, collision induced losses, hidden-node losses and of distinguishing these losses from the unfairness caused by exposed nodes and capture effects.

\section{Acknowledgements}
Many thanks to Ken Duffy and Giuseppe Bianchi for their helpful comments.

\bibliographystyle{IEEEtran}
\bibliography{IEEEabrv,interference_manuscript}

\begin{thebibliography}{10}
\providecommand{\url}[1]{#1}
\csname url@samestyle\endcsname
\providecommand{\newblock}{\relax}
\providecommand{\bibinfo}[2]{#2}
\providecommand{\BIBentrySTDinterwordspacing}{\spaceskip=0pt\relax}
\providecommand{\BIBentryALTinterwordstretchfactor}{4}
\providecommand{\BIBentryALTinterwordspacing}{\spaceskip=\fontdimen2\font plus
\BIBentryALTinterwordstretchfactor\fontdimen3\font minus
  \fontdimen4\font\relax}
\providecommand{\BIBforeignlanguage}[2]{{%
\expandafter\ifx\csname l@#1\endcsname\relax
\typeout{** WARNING: IEEEtran.bst: No hyphenation pattern has been}%
\typeout{** loaded for the language `#1'. Using the pattern for}%
\typeout{** the default language instead.}%
\else
\language=\csname l@#1\endcsname
\fi
#2}}
\providecommand{\BIBdecl}{\relax}
\BIBdecl

\bibitem{txop_technique}
D.~Giustiniano, D.~Malone, D.~J. Leith, and K.~Papagiannaki, ``{Measuring
  Transmission Opportunities in 802.11 Links},'' \emph{{IEEE/ACM Transactions
  on Networking}}, vol.~18, no.~5, pp. 1516 --1529, Oct. 2010.

\bibitem{measurements}
D.~J. Leith and D.~Malone, ``{Field Measurements of 802.11 Collision, Noise and
  Hidden-Node Loss Rates},'' in \emph{IEEE Proc. 8th International Symp. on
  Modeling and Opt. in Mobile, Ad Hoc and Wireless Networks (WiOpt)}, 2010, pp.
  412--417.

\bibitem{cog_rad1}
{K. Rele and D. Roberson and B. Zhang and L. Li and Y. B. Yap and T. Taher and
  D. Ucci and K. Zdunek}, ``{A Two-Tiered Cognitive Radio System for
  Interference Identification in 2.4 GHz ISM Band},'' in \emph{7th IEEE
  Consumer Commun. and Networking Conf. (CCNC)}, Jan. 2010, pp. 1--5.

\bibitem{qiao}
D.~Qiao, S.~Choi, and K.~Shin, ``{Goodput Analysis and Link Adaptation for IEEE
  802.11a Wireless LANs},'' \emph{{IEEE Trans. on Mobile Computing}}, vol.~1,
  no.~4, pp. 278--292, Oct-Dec. 2002.

\bibitem{Haratcherev}
\BIBentryALTinterwordspacing
I.~Haratcherev, K.~Langendoen, R.~Lagendijk, and H.~Sips, ``Hybrid rate control
  for ieee 802.11,'' in \emph{Proceedings of the 2nd Int. Workshop on Mobility
  Management \& Wireless Access Protocols (MobiWac)}, ser. MobiWac '04.\hskip
  1em plus 0.5em minus 0.4em\relax New York, NY, USA: ACM, 2004, pp. 10--18.
  [Online]. Available: \url{http://doi.acm.org/10.1145/1023783.1023787}
\BIBentrySTDinterwordspacing

\bibitem{Zhou}
{T. Zhou and X. Wang and W. Hou}, ``{A Fast Collision Detection Algorithm in
  IEEE 802.11 through Physical Layer SINR Monitoring},'' in \emph{{Proc. of the
  73rd IEEE Vehic. Tech. Conf. (VTC)}}, May {2011}.

\bibitem{Aguayo}
D.~Aguayo, J.~Bicket, S.~Biswas, G.~Judd, and R.~Morris, ``Link-level
  measurements from an 802.11b mesh network,'' in \emph{Proc. of the 2004 Conf.
  on Applications, Tech., Architectures, and Protocols for Computer Commun.
  (SIGCOMM)}, 2004, pp. 121--132.

\bibitem{Rayanchu}
{S. Rayanchu and A. Mishra and D. Agrawal and S. Saha and S. Banerjee},
  ``{Diagnosing Wireless Packet Losses in 802.11: Separating Collision from
  Weak Signal},'' in \emph{{27st Annual IEEE Int. Conf. on Computer Commun.
  (INFOCOMM)}}, Apr. {2008}, pp. 735--743.

\bibitem{cog_rad0}
J.~M. III and G.~Q.~M. Jr., ``{Cognitive radio: making software radios more
  personal},'' \emph{{IEEE Personal Communications}}, vol.~6, no.~4, pp.
  13--18, Aug. 1999.

\bibitem{ARF}
A.~Kamerman and L.~Monteban, ``Goodput analysis and link adaptation for ieee
  802.11a wireless lans,'' \emph{{Bell Labs Tech. Jour.}}, pp. 118--133, Summer
  1997.

\bibitem{RBAR}
{G. Holland and N. Vaidya and P. Bahl}, ``{A Rate-Adaptive MAC Protocol for
  Multi-Hop Wireless Networks},'' in \emph{{Proc. of the Int. Conf. on Mobile
  Computing and Networking (MobiCom)}}, July {2001}, pp. 236--251.

\bibitem{RRAA}
\BIBentryALTinterwordspacing
S.~H.~Y. Wong, H.~Yang, S.~Lu, and V.~Bharghavan, ``Robust rate adaptation for
  802.11 wireless networks,'' in \emph{{Proc. of the 12th Annual Int. Conf. on
  Mobile Computing and Networking (MobiCom)}}.\hskip 1em plus 0.5em minus
  0.4em\relax New York, NY, USA: ACM, 2006, pp. 146--157. [Online]. Available:
  \url{http://doi.acm.org/10.1145/1161089.1161107}
\BIBentrySTDinterwordspacing

\bibitem{Pang}
Q.~Pang, V.~C.~M. Leung, and S.~C. Liew, ``A rate adaptation algorithm for ieee
  802.11 wlans based on mac-layer loss differentiation,'' in \emph{2nd Int.
  Conf. on Broadband Networks (BroadNets)}, Oct. 2005, pp. 659 -- 667.

\bibitem{cara}
{J. Kim and S. Kim and S. Choi and D. Qiao}, ``{CARA: Collision-aware Rate
  Adaptation for IEEE 802.11 WLANs},'' in \emph{{25th Annual IEEE Int. Conf. on
  Computer Commun. (INFOCOMM)}}, Apr. 2006.

\bibitem{dyspan}
D.~Malone, P.~Clifford, D.~Reid, and D.~J. Leith, ``{Experimental
  Implementation of Optimal WLAN Channel Selection Without Communication},'' in
  \emph{Proc. IEEE Dynamic Spectrum Access Netw. (DySPAN)}, 2007, pp. 316--319.

\bibitem{PPER_smartrate}
{M. Khan and D. Veitch}, ``{Isolating Physical PER for Smart Rate Selection in
  802.11},'' in \emph{{28th Annual IEEE Int. Conf. on Computer Commun.
  (INFOCOMM)}}, Apr. 2009, pp. 1080--1088.

\bibitem{error_id_Zakhor}
M.~N. Krishnan, S.~Pollin, and A.~Zakhor, ``Local estimation of probabilities
  of direct and staggered collisions in 802.11 wlans,'' in \emph{IEEE Global
  Telecommunications Conference, 2009. (GLOBECOM)}, Dec. 2009, pp. 1--8.

\bibitem{Kim}
{E.-I. Kim and J.-R. Lee and D.-H. Cho}, ``{Throughput Analysis of Data Link
  Protocol with Adaptive Frame Length in Wireless Networks},'' \emph{{AEU Int.
  J. Electron. Commun.}}, pp. 1--8, 2003.

\bibitem{pktlen-adapt1}
J.~Yin, X.~Wang, and D.~P. Agrawal, ``{Optimal packet size in error-prone
  channel for IEEE 802.11 distributed coordination function},'' in \emph{IEEE
  Wireless Communications and Networking Conference (WCNC)}, vol.~3, March
  2004, pp. 1654--1659.

\bibitem{pktlen-adapt2}
{X. He and F. Y. Li and J. Lin}, ``{Link Adaptation with Combined Optimal Frame
  Size and Rate Selection in Error-prone 802.11n Networks},'' in \emph{{Proc.
  of IEEE Int. Symp. on Wireless Commun. Systems (ISWCS)}}, {2008}, pp.
  733--737.

\bibitem{pktlen-adapt3}
{S. Choi and K. G. Shin}, ``{A Class of Adaptive Hybrid ARQ Schemes for
  Wireless Links},'' \emph{{IEEE Trans. on Vehic. Tech.}}, pp. 777--790, 2001.

\bibitem{pktlen-adapt4}
{S. Ci and H. Sharif}, ``{An Link Adaptation Scheme for Improving Throughput in
  the IEEE 802.11 Wireless LAN},'' in \emph{{Proc. of 27th Annual IEEE Conf. on
  Local Computer Networks (LCN)}}, Nov. {2002}, pp. 205--208.

\bibitem{pktlen-adapt5}
P.~Lettieri and M.~B. Srivastava, ``Adaptive frame length control for improving
  wireless link throughput, range, and energy efficiency,'' in \emph{Proc. of
  the IEEE 17th Annual Joint Conf. of the IEEE Computer and Commun. Soc.
  (INFOCOM)}, vol.~2, Mar 1998, pp. 564--571.

\bibitem{Krishnan-infocomm}
{M. Krishnan and E. Haghani and A. Zakhor}, ``{Packet Length Adaptation in
  WLANs with Hidden Nodes and Time-Varying Channels},'' in \emph{{submitted to
  IEEE INFOCOMM}}, 2011.

\bibitem{wolff82}
R.~W. Wolff, ``{Poisson Arrivals See Time Averages},'' \emph{{Operations
  Research}}, vol.~30, no.~2, pp. 223--231, 1982.

\bibitem{melamed90}
B.~Melamed and W.~Whitt, ``{On Arrivals That See Time Averages},''
  \emph{{Operations Research}}, vol.~38, no.~1, pp. 156--172, 1990.

\bibitem{baccelli2009}
F.~Baccelli, S.~Machiraju, D.~Veitch, and J.~Bolot, ``{The Role of PASTA in
  Network Measurement},'' \emph{{IEEE/ACM Trans. on Networking}}, vol.~17,
  no.~4, pp. 1340--1353, 2009.

\bibitem{nonsat}
{D. Malone and K. Duffy and D.J. Leith}, ``{Modeling the 802.11 Distributed
  Coordination Function in Nonsaturated Heterogeneous Conditions},''
  \emph{{IEEE/ACM Trans. on Networking}}, vol.~15, no.~1, pp. 159 --172, Feb.
  2007.

\bibitem{prop_soln}
\BIBentryALTinterwordspacing
I.~Tinnirello, D.~Giustiniano, L.~Scalia, and G.~Bianchi, ``{On the
  side-effects of proprietary solutions for fading and interference mitigation
  in IEEE 802.11b/g outdoor links},'' \emph{{Computer Networks}}, vol.~53,
  no.~2, pp. 141 -- 152, 2009, qoS Aspects in Next-Generation Networks.
  [Online]. Available:
  \url{http://www.sciencedirect.com/science/article/B6VRG-4TTMK0V-1/2/71fc79de%
16b0855ac93813d9babf0dc1}
\BIBentrySTDinterwordspacing

\bibitem{validation}
K.~D. Huang, K.~R. Duffy, and D.~Malone, ``{On the validity of IEEE 802.11 MAC
  modeling hypotheses},'' \emph{{IEEE/ACM Transactions on Networking}},
  vol.~18, no.~6, pp. 1935 -- 1948, 2010.

\bibitem{mwo_int}
A.~Kamerman and N.~Erkocevic, ``{Microwave Oven Interference on Wireless LANs
  Operating in the 2.4 GHz ISM Band},'' in \emph{IEEE Int. Symp. on Personal,
  Indoor and Mobile Radio Commun. (PIMRC)}, vol.~3, September 1997, pp.
  1221--1227.

\bibitem{Anderson2008Commodity}
{E. Anderson and G. Y. and C. Phillips and D. Sicker and D. Grunwald},
  ``{Commodity AR52XX-Based Wireless Adapters as a Research Platform},''
  {University of Colorado at Boulder}, Department of Computer Science, Campus
  Box 430, Technical Report CU-CS-XXXX-08, April 2008.

\bibitem{ge_channel}
M.~Mushkin and I.~Bar-David, ``{Capacity and coding for the Gilbert-Elliot
  channels},'' \emph{{IEEE Transactions on Information Theory}}, vol.~35,
  no.~6, pp. 1277--1290, Nov. 1989.

\bibitem{ge_1}
S.~Sivaprakasam and K.~S. Shanmugan, ``{An equivalent Markov model for burst
  errors in digital channels},'' \emph{{IEEE Transactions on Communications}},
  vol.~43, no. 234, pp. 1347--1355, Feb/Mar/Apr 1995.

\bibitem{ge_3}
{M. Yajnik and S. Moon and J. Kurose and D. Towsley}, ``{Measurement and
  Modeling of the Temporal Dependence in Packet Loss},'' in \emph{{18th Annual
  IEEE Int. Conf. on Computer Commun. (INFOCOMM)}}, vol.~1, Mar. 1999, pp. 345
  -- 352.

\bibitem{ge_2}
H.~S. Wang and N.~Moayeri, ``{Finite-state Markov channel-a useful model for
  radio communication channels},'' \emph{{IEEE Transactions on Vehicular
  Technology}}, vol.~44, no.~1, pp. 163--171, Feb. 1995.

\end{thebibliography}

\begin{IEEEbiography}{Brad Zarikoff} received the B.Eng. degree with distinction in electrical engineering from the University of Victoria, Victoria, Canada, in 2002 and the M.A.Sc. and Ph.D. degrees from Simon Fraser University, Burnaby, Canada, in 2004 and 2008, respectively. He is currently a research fellow at the Hamilton Institute, National University of Ireland Maynooth. His current research interests include interference mitigation, power line communication networks, and synchronisation for network MIMO systems.
\end{IEEEbiography}

\begin{IEEEbiography}{Doug Leith} graduated from the University of Glasgow in 1986 and was awarded his PhD, also from the University of Glasgow, in 1989. In 2001, Prof. Leith moved to the National University of Ireland, Maynooth to assume the position of SFI Principal Investigator and to establish the Hamilton Institute (www.hamilton.ie) of which he is Director.  His current research interests  include the analysis and design of network congestion control and  resource allocation in wireless networks.
\end{IEEEbiography}

\end{document}